# Citation Cascade and the Evolution of Topic Relevance

Chao Min†
School of Information Management, Nanjing University, #163 Xianlin Avenue, Nanjing, Jiangsu 210023, China.
E-mail: mc@nju.edu.cn

Qingyu Chen†
School of Computing and Information Systems, The University of Melbourne, Parkville VIC 3052, Australia.
Current affiliation: National Center for Biotechnology Information, National Library of Medicine, National Institutes of Health, 8600 Rockville Pike, Bethesda, MD 20894, USA.
E-mail: qingyu.chen@unimelb.edu.au

Erjia Yan
College of Computing and Informatics, Drexel University, Philadelphia, PA 19104, U.S.A.
E-mail: ey86@drexel.edu

Yi Bu
Department of Information Management, Peking University, Beijing 100871, China
Luddy School of Informatics, Computing, and Engineering, Indiana University, Bloomington, IN 47408, U.S.A.
E-mail: buyi@pku.edu.cn

Jianjun Sun
School of Information Management, Nanjing University, #163 Xianlin Avenue, Nanjing, Jiangsu 210023, China.
E-mail: sjj@nju.edu.cn

## Abstract

Citation analysis, as a tool for quantitative studies of science, has long emphasized direct citation relations, leaving indirect or high-order citations overlooked. However, a series of early and recent studies demonstrate the existence of indirect and continuous citation impact across generations. Adding to the literature on high-order citations, we introduce the concept of a *citation cascade*: the constitution of a series of subsequent citing events initiated by a certain publication. We investigate this citation structure by analyzing more than 450,000 articles and over 6 million citation relations. We show that citation impact exists not only within the three generations documented in prior research, but also in much further generations. Still, our experimental results indicate that two to four generations are generally adequate to trace a work's scientific impact. We also explore specific structural properties—such as depth, width, structural virality, and size—which account for differences among individual citation cascades. Finally, we find evidence that it is more important for a scientific work to inspire trans-domain (or indirectly related domain) works than to receive only intra-domain recognition in order to achieve high impact. Our methods and findings can serve as a new tool for scientific evaluation and the modeling of scientific history.

**Keywords**

Citation generation, indirect citation, high-order citation, citation cascade, citation diffusion

# Introduction

The issue of indirect (or high-order) citation is a long-standing problem in citation analysis research. As far as we know, Rousseau (1987) was among the earliest researchers to take into account the indirect influence (that is, citations to citations, or references of references) of a scientific publication. This perspective bears important implications for many applications of citation analysis, such as scientific evaluation, information search and retrieval, and the modeling of the history of science (Zunde, 1971). Researchers have further explored this topic from both theoretical (Fragkiadaki & Evangelidis, 2014) and empirical (Atallah & Rodriguez, 2006; Fragkiadaki et al., 2011; Huang et al., 2018; Liu, Lu, & Ho, 2012) vantage points. Especially notable in this connection are a recent series of studies by Hu and Rousseau (2016; 2017; Hu, Rousseau & Chen, 2011). Yet most existing efforts are either case studies or based on small datasets, and the citations under consideration are usually limited to three generations. Consequently, we still lack a systematic understanding of the basic concepts and properties of indirect citations, especially those in "deep" citation space. The present study aims to ameliorate this situation, revealing empirical details and characteristics of indirect citations through a systematic investigation.

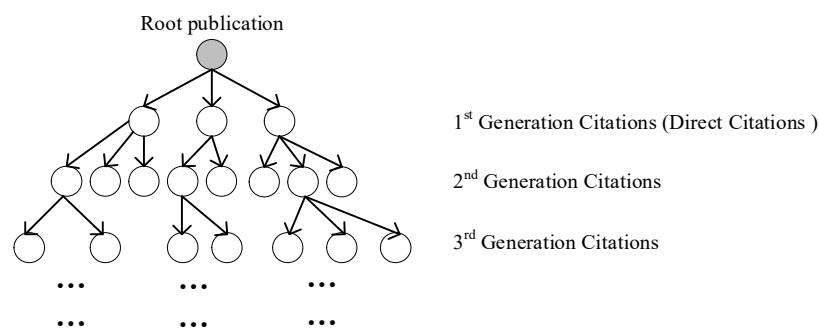

Figure 1. A simplified citation cascade schematic. The arrow means "is cited by" (and not

"cites"). That is, the arrow follows the knowledge flow[1].

While direct citations of scientific literature have long been a central concern in theoretical and applied bibliometric research, indirect citations have been relatively neglected. "Direct citations" here refers to first-order citations to a scientific output (usually a publication), which implies its direct impact; "indirect citations" are higher-order citations, which provide evidence of indirect impact. Kostoff (1998) argued that citations could "serve as a 'radioactive tracer' of research impacts" (p. 29), an application he judged to be "very fruitful" (p. 30) but as yet underdeveloped. To a scholar—or even a user—of social networks, the "radioactive tracer" metaphor bears a clear resemblance to the information cascading process in such platforms as Facebook (Cheng, Adamic & Kleinberg et al., 2016), Twitter (Goel, Anderson & Hofman et al., 2015), WeChat (Liu, Qu & Chen et al., 2017) and Weibo (Huang & Sun, 2014). "Radioactive" is also an apt description for the expanding citation process (Chen, 2018; Huang, 2018) that we (Min, Sun & Ding, 2017) term the *citation cascade* (Figure 1). Hu and Rousseau's recent introduction (2016; 2017) of "under-cited influential publications" manifests the importance of taking indirect citations into consideration: scientific contributions, as these scholars show, are not always visible from a direct citation perspective. In fact, as early as 1987, Rousseau posed the problem of high-order citations and their potential value for research on the rules of scientific citation and the history of science. Since that time, however, further research on indirect citations seems to have been impeded by historical limitations such as accumulation of bibliometric data, construction of citation databases, and lack of computing power.

As a result, the literature on quantitative studies of science has extensively addressed the direct citation influence of scientific works (Chavalarias & Cointet, 2013; Dietz, Bickel & Scheffer, 2007; Mazloumian et al., 2013; Sinatra, 2016), while indirect or high-order citation impact has been underexplored. We contribute to the literature on citation analysis and probe into the area of deep citation space (> 450,000 articles) by investigating the structural properties of citation diffusion as well as the evolutionary effects of research relevance on citation structure. To explore the structural expansion of citations within complex networks, we introduced the concept of the citation cascade (Min, Sun & Ding, 2017) and suggested metrics by which to quantify and measure it. We now analyze how citations behave when they reach into a currently unknown depth (e.g., 30 steps from the original publication). Is behavior at this depth totally random, or might it still be somewhat associated with the original publication? The answer is sought first by examining structural properties such as the depth, width, size, and virality of the citation cascade; then by exploring how research topic relevance evolves along the path of citation diffusion; and finally by analyzing the relationship between citation cascades and direct impact.

From a history-of-science standpoint, our analysis provides a new dimension for researchers who are looking to quantify the historical development and evolution of science. From a policy and managerial standpoint, our findings provide empirical support for scientometric practitioners and decision makers who seek to incorporate

---

[1] Please note that a citation cascade is only partly chronological, as a publication in an early citation generation can easily be younger than a publication in a late citation generation.

both direct and indirect citations when evaluating scientific outputs.

# Related work

### From information cascades to citation cascades

*Citation cascade* is a relatively new term (Min, Sun & Ding, 2017); to the best of our knowledge, it was first mentioned in the work of Mazloumian et al. (2011), where it referred to citation bursts experienced by the works of Nobel laureates. Here, we use the phrase somewhat differently to suggest an analogy to the *information cascades* extensively studied by computer scientists. The literature on information cascades addresses the mechanism by which information reaches its audience within an online social system such as Facebook, Twitter, or Instagram. Guille et al. (2013) supply a more general, graph-theoretic definition of a *spreading cascade*: "a directed tree having as a root the first node of the activation sequence" (p. 21). They add that "the tree captures the influence between nodes and unfolds in the same order as the activation sequence" (p. 21). Related studies have focused on quantifying and modeling information cascades on various online platforms, seeking thereby to understand the mechanisms of viral ideas and products. Using a large sample of photo reshare cascades on Facebook, Cheng, Adamic & Dow et al. (2014) found strong evidence for predicting the future growth of a cascade based on five classes of cascade features. In 2016, Cheng et al. further explored Facebook cascades by providing a model for characterizing and predicting cascade recurrence; they found that recurrence is both widespread and predictable for large cascades. Goel, Anderson & Hofman et al. (2015) applied structural measures to study the diffusion of nearly a billion Twitter posts (including news stories, videos, pictures and petitions) and made comparisons between empirical observations and simulation results. Anderson, Huttenlocher & Kleinberg et al. (2015) investigated invitation cascades on the professional-networking site LinkedIn, finding that both structural patterns and temporal growth of LinkedIn cascades are qualitatively different from other types of online diffusion.

Citation cascades operate similarly. A paper initiates a citation cascade when it is cited by papers that in turn are cited by their own successors, and so forth (Min, Sun & Ding, 2017; Chen, 2018). The analogy is not a perfect one: a citation cascade is in some ways qualitatively different from an information cascade. Whereas the same content is spread in a cascade of Facebook posts or Twitter retweets, a citation cascade does not simply disseminate information about the source paper. In addition, information mutation (Kuhn, Perc & Helbing, 2014) and decay (Liu & Kuan, 2016) are more evident in citation cascades, as a citing paper usually draws part of the knowledge from a cited paper and adds knowledge of its own. Mutation and decay effects can be even more significant after several generations of diffusion. Furthermore, in citation networks, one node can, and usually does, have multiple parent nodes (references); in Facebook or Twitter repost networks, every node has at most one parent node. In this respect, the diffusion and interactions concerning information in citation cascades are far more complex than those in online information cascades.

Main-path analysis (Hummon & Doreian, 1989) provides a powerful tool for tackling the problem of diffusion complexity in citation networks. It is a method capable of extracting influential citation relations and thus selecting significant papers in a complicated citation network, though some important papers may be ignored, and knowledge decay is usually not considered (Liu & Kuan, 2016).

**Indirect citation and citation generation**

Whereas literature on main-path analysis considers only a small proportion of representative papers, some other citation-related studies have taken indirect citations into consideration. Rousseau (1987) was among the very first to include both direct and indirect citations (as well as direct and indirect references) in a mathematical technique devised to measure the total influence of citations. Atallah and Rodriguez (2006) took advantage of citation chains to measure the quality of a patent by calculating cumulative citations (both direct and indirect), with each citation arithmetically weighted according to its distance to the original patent. Fragkiadaki et al. (2011) and Hu, Rousseau & Chen (2011) proposed their own indicators that take both direct and indirect citations into account to measure total influence. Liu, Lu, and Ho (2012) instead used indirect citations to a researcher's works to determine his or her overall influence, as well as the level of those works' association with the mainstream in a scientific field. A more recent development in indirect citation research is Hu & Rousseau's (2016; 2017) introduction of the notion of "under-cited influential publications." In these cases, the publications themselves contain foundational scientific discoveries that attract fewer citations than expected but trigger more influential research; thus, their influence is shown by the presence of significant follow-up research. Using up to three generations of citations, Hu & Rousseau showed that scientific contributions are not always visible from a direct-citation perspective and argued that taking more than one citation generation into account may help us recognize a work's true value.

A related issue for researchers is that of citation generation (Fragkiadaki & Evangelidis, 2014), the "order" indicated in the phrase "high-order citation" (Sizov & Bahn, 2017). Fragkiadaki & Evangelidis (2014) defined a generation of citations as "the collection of papers that cite a target paper either directly (first generation) or indirectly (via a path in the citation graph originating from a source paper and ending to the target paper)" (p. 272). The concepts of backward and forward citation generations were originally presented by Rousseau (1987). In brief, backward citations are a collection of papers referenced by a paper of interest, along with their references, and so forth; forward citations, the variety more typically discussed, are a collection of papers citing a paper of interest, along with the papers citing *those* papers, and so on. Hu, Rousseau & Chen (2011) formalized these definitions and provided methods for deriving indicators based on different citation generations. They (Hu et al., 2011) further pointed out that such definitions will vary depending on whether a paper can be counted more than once in a generation and whether a paper can recur in different generations, since there might be different paths leading from a source paper to a target paper. Dervos,

Samaras & Evangelidis et al. (2006) proposed the cascading citation indexing framework (c$^2$IF) for assessing publications' impact in promoting science and technology and introduced the concept of the chord in citation generations.

The number of citation generations to be included in any given study remains an important but unresolved issue in the research community (Liu & Kuan, 2016). Rousseau (1987) deemed two to four citation generations reasonable, while Kosmulski (2010) considered up to the second generation; Hu et al. (2011), as well as Fragkiadaki et al. (2011), drew the line at three generations. In general, however, it seems that few efforts have been made to explore the citation space further than four generations.

## Research relevance and citation networks

A significant difference between information cascades and citation cascades lies in the relevance of the source node to subsequent generations. While the core information content remains almost unchanged in such information cascades as Twitter retweets, the original knowledge contained in a source paper generally assumes a weaker relevance as one proceeds further along a citation pathway. This pattern coincides with many general phenomena in social networks (Travers & Milgram, 1969; Watts, 2004; Christakis & Fowler, 2009), including people's emotions, intimate relationships, health, economics, and politics. Travers & Milgram's (1969) experiments with "six degrees of separation" indicate that a person can be connected to almost everybody else in the world within six degrees through his or her social network. Christakis & Fowler (2009) further proposed the theory of "three degrees of impact," suggesting that a person's social impact is confined to three degrees of connectedness. They reason that connections within three degrees are strong links that affect behavior, whereas connections beyond three degrees are weak links that can only deliver information.

It is interesting to investigate whether similar phenomena exist in citation networks, which can also be considered as "social networks" of academic publications. Both Elmacioglu & Lee's (2005) study of the DBLP database and Nascimento, Sander & Pound's (2003) examination of the SIGMOD network showed characteristics of small-world networks, with the average path lengths stabilizing at approximately six. Radev, Joseph & Gibson et al. (2016), in their investigation of ACL Anthology data, also found that the paper citation network in computational linguistics has an average directed shortest path length less than six.

Some studies have been conducted to quantify the relevance of such links. Dietz, Bickel & Scheffer (2007) defined the strength of influence as the similarity between the topic mixtures of the cited and citing papers. Using features such as similarity between abstracts, Valenzuela, Ha, & Etzioni (2015) proposed a supervised classification method to identify meaningful citations. Kim, Baek & Song (2017) utilized references and keywords of cited and citing papers to measure their relevance, based on which they constructed weighted citation networks to trace topic diffusion in biomedical literature. Zhu, Turney & Lemire et al. (2015) proved that a high degree of semantic similarity between a cited and a citing paper indicates that the former has a significant impact on the latter. These works provide practical references for analyzing research relevance along citation chains.

# Method and Data

## Definition and quantification of citation cascade

### Definition of citation cascade

Referencing Guille, Hacid & Favre's (2013) definition of information cascade, we define the *citation cascade* of a publication as an acyclic graph having as a root the root publication, as children the direct and indirect citing publications, and as edges the direction of knowledge flow among these publications. We further define the following related concepts.

**Root**: the root publication, considered as the research object.

**Direct citing publications**: Publications that directly cite the root.

**Indirect citing publications**: Child publications in the cascade other than direct citing publications.

**$N^{th}$ order citations ($N^{th}$ generation citations)**: Publications that are N steps from the root. For instance, a direct citation is a $1^{st}$ order citation (or $1^{st}$ generation citation), and a citation to a $1^{st}$ order publication is a $2^{nd}$ order citation (or $2^{nd}$ generation citation).

**Direction of knowledge flow**: the direction of knowledge flow is from cited publication to citing publication. Note that the direction of knowledge flow is opposite the citing direction; e.g., if A is cited by B, then citing direction is B→A, yet knowledge flow is A→B.

In a citation cascade, there can be multiple paths leading from the root to the same child node. That is, the child node could simultaneously be the root's $N^{th}$ order citation, $(N-1)^{th}$ order citation, $(N-2)^{th}$ order citation, and so forth. Nevertheless, the length of the shortest path between the root and a child node is certain to be unique. The possibility that there could be multiple paths between two nodes not only makes it possible for the same child node to belong to different citation generations via different paths, but also results in the possibility that a child node in a certain citation generation might stem from the root via multiple paths. Therefore, we have four possible definitions of $N^{th}$ order citation, based on the determination of the following two criteria (Hu, Rousseau & Chen, 2011; Fragkiadaki & Evangelidis, 2014):

**Criterion 1**: Is a publication allowed to appear simultaneously in different citation generations?

**Criterion 2**: Can a publication appear repeatedly in the same citation generation?

Here, to avoid complicating the problem, we allow a publication both to appear simultaneously in different citation generations (Criterion 1) and to appear repeatedly in the same citation generation (Criterion 2). The reason is that from the perspective of information flows, different paths bear different traffic and thus should be treated separately. This is a relatively simple and clear definition.

## Measurement of citation cascade

**(1) Depth**
　　A child node's *depth* is the shortest path length from the root to that node within the citation cascade. The maximum value of any child node's depth is the depth of the citation cascade, which reflects how deep the citation cascade can reach.

**(2) Width**
　　The number of nodes in a citation generation is called the generation's *width*. The maximum width of any generation is taken as the width of the citation cascade, which reflects how wide the citation cascade can extend.

**(3) Structural virality**
　　*Structural virality* is a concept proposed by Goel et al. (2015) to measure a cascade's depth and width in a single composite indicator. Goel et al. (2015) suggest a list of candidate indicators, among which we choose the average depth of all nodes:

$$Average\ depth\ =\ \frac{1}{|T|-1}\sum_{v\in T, v\neq r} d(v,r)$$

where $T$ is the citation cascade initiated by publication $r$, and $d(v,r)$ is the shortest path length between nodes $v$ and $r$.

**(4) Size**
　　The number of nodes in a citation cascade is the *size* of the citation cascade. The size dramatically increases as the depth of a citation cascade increases.

**(5) Topic relevance**
　　A problem to note when the size of a citation cascade rapidly increases is that the newly added nodes might have little to do with the root publication. That is, when N exceeds a certain range, the $N^{th}$ order citation might join the cascade for reasons other than the impact of the root publication and thus have a rather weak relation with the root. Therefore, we introduce the concept of *diffusion impact*, which refers to how much the child nodes (direct or indirect citations) are influenced by the root publication in joining the citation cascade. Although some efforts (Dietz, Bickel & Scheffer, 2007) have been made to accurately measure how much influence the cited publication has on the citing publication's citing behavior, further study is still needed to fulfill this goal. We assume here that the more correlated the cited publication and the citing publication, the more influence the former has on the latter (Zhu et al., 2015), making topic relevance between citing and cited publications a proxy for diffusion impact. This is, we admit, an imperfect substitution, but relevance can indeed reflect the impact of the cited publication in some ways. For our dataset, introduced more thoroughly below, we used the Jaccard similarity between two articles' PACS (Physics and Astronomy Classification Scheme) codes to measure their topic relevance:

$$R = \frac{|P_i \cap P_j|}{|P_i \cup P_j|}$$

Where $P_i$ and $P_j$ are the second-level PACS code sets of articles *i* and *j* respectively.

## Data and processing

The American Physical Society (APS) dataset covers a series of physics journals. The 2013 version of the dataset includes more than 450,000 articles, with over 6 million citations among them. Moreover, every article published from 1975 to 2009 was assigned one or more PACS codes to indicate its research topics. Although PACS has some known limitations (Smith, 2019), we adopt this subject scheme in calculations of topic relevance in light of its efficiency and effectiveness (Jia, Wang & Szymanski, 2017). Based on APS citation relations, we construct a citation cascade for every article in the dataset. For every cascade whose root was published during 1975–2009, we further calculate depth, cascade width, width of $N^{th}$ order citation, structural virality, size, and topic relevance between the root article and each child node. In addition, we calculate the mean, median and variance of topic relevance for all nodes in the same citation generation. In particular, the topic relevance of each article in the $1^{st}$ citation generation is recorded for follow-on analysis[2].

To identify typical patterns from various types of citation curves, we apply a time series clustering technique[3]. This technique first normalizes the time series data to a suitable scale, then clusters the data so that similar time series are grouped into the same category while dissimilar ones are separated.

## Structural characteristics of citation cascades

### Depth and structural virality

Among all cascades in the APS dataset, half have a depth less than 10, while the deepest cascade reaches more than 60 generations. About 20% of the cascades are depth 20 or greater; the percentage decreases to 10% for cascades whose depth is greater than 30. Very deep cascades—those with depth > 40—are rare.

Similarly, across the entire dataset, the maximum structural virality value is greater than 50, yet the value for most citation cascades (~90%) is less than 15. To give a visual sense of the cascades' structures, Figure 2 presents plots of cascades with viralities approximately 2, 5, 10, and 15, respectively.

---

[2] The codes for this study are available at https://github.com/qingyu-qc/network_analytics.
[3] We use tslearn, a Python package that provides machine learning tools for the analysis of time series, available here https://github.com/rtavenar/tslearn.

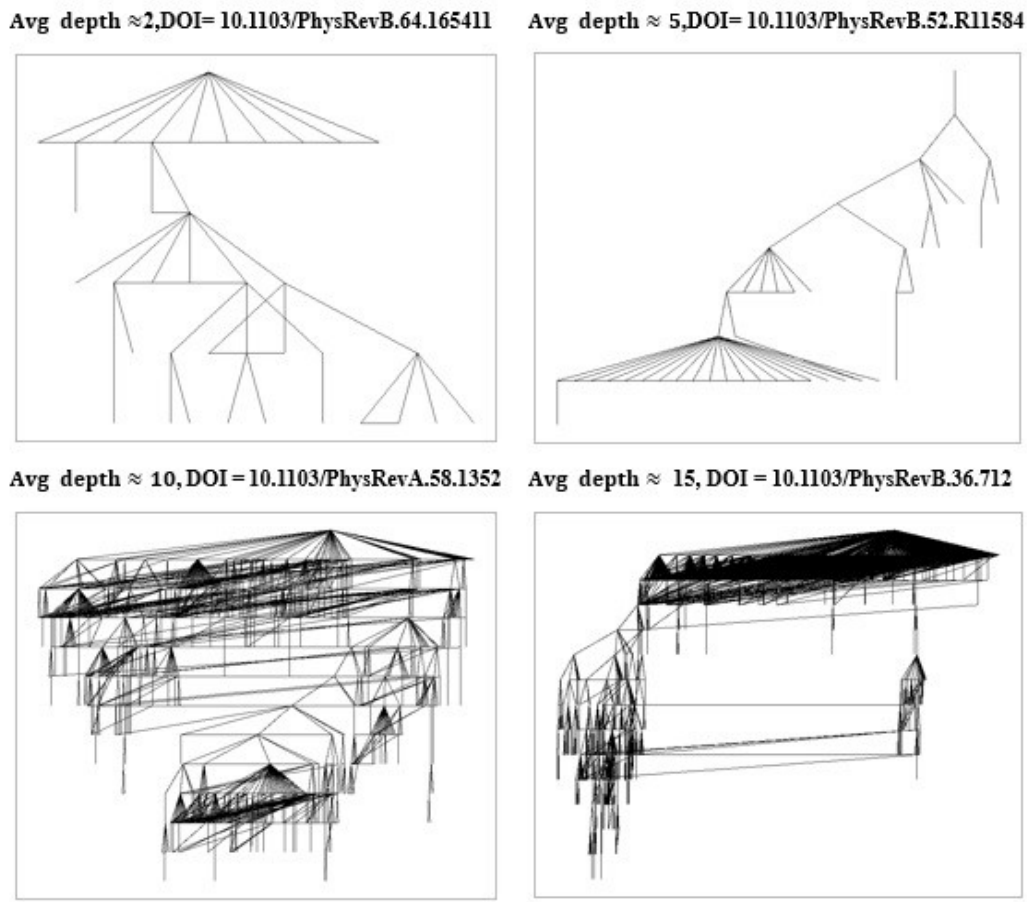

Figure 2. Citation cascades with structural virality ≈ 2, 5, 10, 15

## Generation width

As defined above, the number of nodes in a citation generation is the width of that generation. How wide do citation generations get along a given citation cascade? To answer this question, we construct a width distribution curve by setting citation generation number as the x-axis and the corresponding width as the y-axis. This curve shows vividly how the width of a citation generation changes as generations deepen. We find, however, that the width distribution curve varies greatly between different cascades. To identify typical patterns from tens of thousands of width distribution curves, we apply a clustering technique via the Python package tslearn[4] (described in Method) so that curves with similar shapes can be clustered into the same category. We use the K-means clustering algorithm integrated in the package and try different K values to determine the proper values to reflect the number of potential categories. To obtain a representative sample of the data, we select cascades with a depth of 10, 20, 30, and 50 respectively. In doing so, it quickly becomes evident that cascades with depth of 10, 20, and 30 are very similar in terms of shape, while cascades with depth of 50 are markedly different. We therefore show only the results for cascades with depth 10

---

[4] https://github.com/rtavenar/tslearn

and 50.

There are 7,504, 9,215, and 1,409 cascades with depths of 10, 20, and 30 respectively. At each of these depths, the generation width shows a general increase-and-then-decrease pattern, typically with a single peak on the distribution curve. For cascades of depth 10 (Figure 3), the peak can appear as early as the third or fourth generation, or as late as the sixth or seventh. In addition, these peaks may be either sharp or mild. Interestingly, the increase-and-then-decrease trend is not absolute: in some cases, a second peak appears after the first decrease. This indicates that for an average citation cascade with these depths, the number of papers in a citation generation first increases and then decreases. The middle citation generations always have a relatively large number of papers.

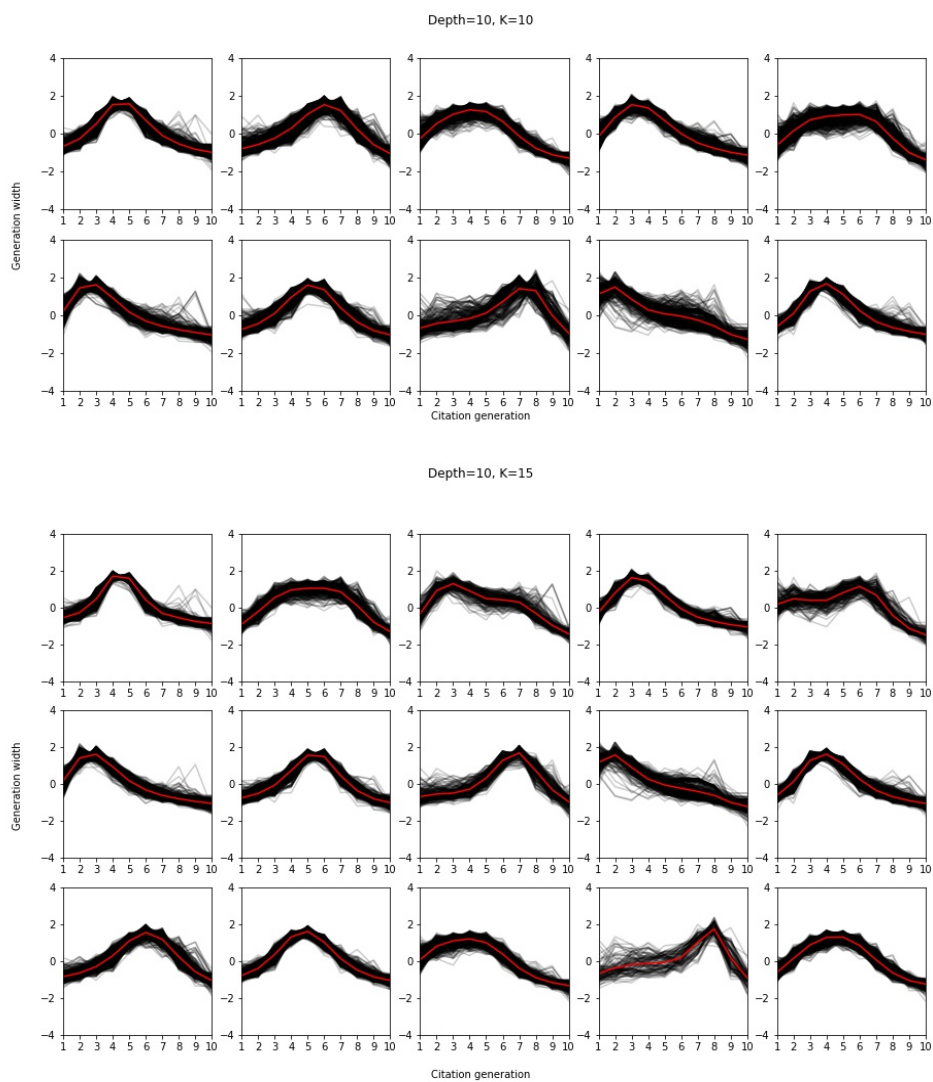

Figure 3. Typical patterns of width distribution curves at depth = 10 (7,504 cascades, K = 10, 15)

Citation cascades with a depth of 50 (Figure 4), however, exhibit a different shape of width distribution. Only 206 papers in the dataset reach such depth. The width of their successive citation generations fluctuates more dramatically, forming a two-stage separation in most cases; this separation appears near the $20^{th}$ citation generation, dividing these cascades into two subsequent increase-then-decrease phases that are clearly independent of each other. This almost universal pattern indicates that the linkage from the root publication can extend to about 20 generations (the first phase), after which the continuation of the cascade might be driven by papers on other topics (the second phase).

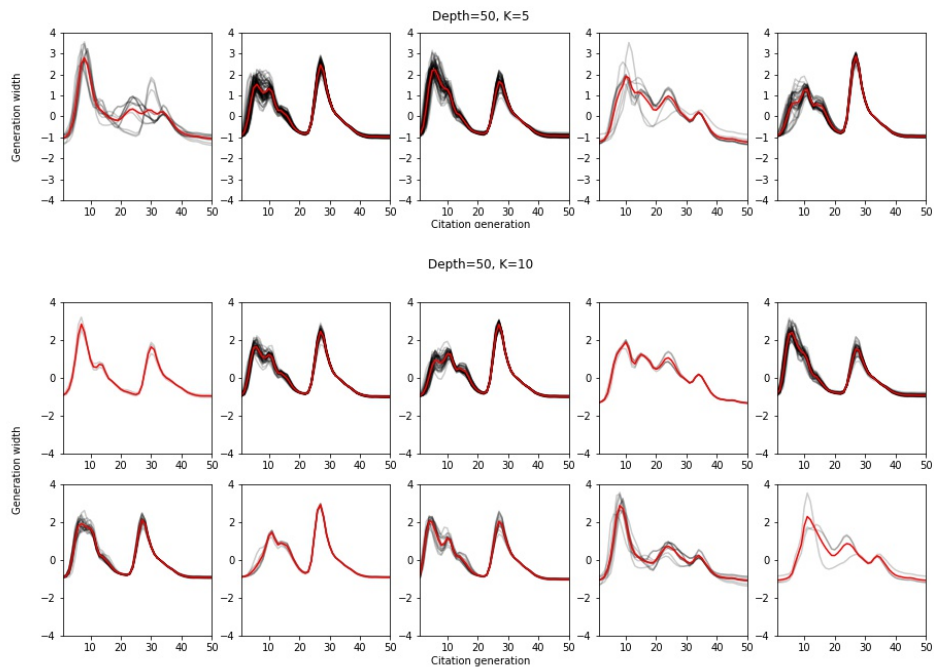

Figure 4. Typical patterns of width distribution curves at depth = 50 (206 cascades, K = 5, 10)[5]

## Cascade size

Figure 5 shows that the number of citation cascades drops rapidly as their size increases. Although the maximal size approaches 250,000, the proportion of cascades larger than 50,000 nodes is less than 20%, indicating that the sizes of most cascades are within a relatively small range. A surprising observation is that the distribution curve of cascade size breaks up into two parts[6] between approximately 150,000 and 220,000. That is, no citation cascade exist within the fault zone, yet they do exist in both larger and smaller sizes. More interestingly, the two split parts exhibit a similar pattern of sharp-decrease-and-then-stabilize, like a phase transition from the former status to the latter.

---

[5] As there are only 206 cascades with depth 50, the K values chosen for clustering is accordingly lowered.
[6] We investigated the cause of the gap and found no mistake in either the code or the raw data. We tend to believe the gap is inherent in the APS dataset, arising from (for instance) the limited number of journals indexed by APS, the later founding dates of some journals, and changes in journal names.

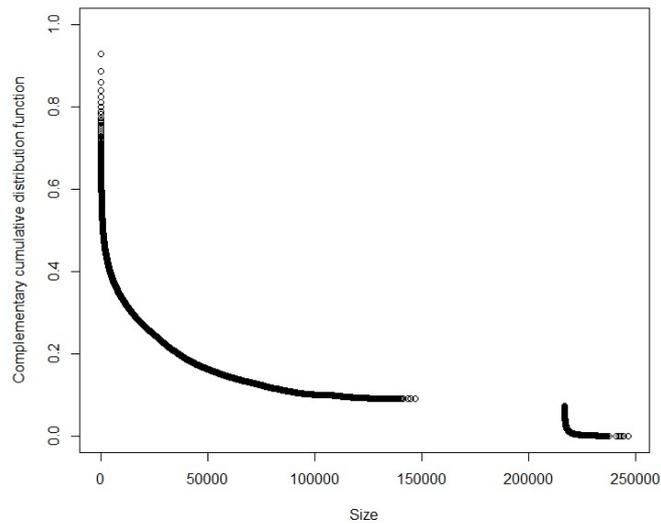

Figure 5. Distribution of the size of citation cascades

## Topic relevance in citation cascades

As stated in the Method section, we use the Jaccard similarity of PACS codes between two articles to measure their topic relevance. Compared with text-based topic modeling, this is simple, handy, and reliable, avoiding the computational problems caused by the rapid growth of citations in cascades.

## Typical relevance curve patterns

Given the root article in a citation cascade, we can measure its relevance to each direct or indirect citing article, as well as its relevance to an entire citation generation. We quantify the latter as the average of the relevance between the root article and each citing article in the same generation. By setting generation number as the x-axis and the average relevance for the generation as the y-axis, we can then obtain a relevance curve that depicts how topic relevance changes with the citation generation of an article. As before, we choose sample cascades with depths 10, 20, 30, and 50 respectively[7], then apply clustering techniques to investigate the typical patterns of relevance curves. In Figures 6 and 7, the x-axis is citation generation number and the y-axis is the normalized[8] average relevance of the generation. Somewhat counterintuitively, the topic relevance does not always monotonically decrease with an increase in citation generation. They do not even display a simple linear relationship; rather, different cascades exhibit distinct trends.

---

[7] There are 7,504, 9,215, 1,409, and 206 citation cascades respectively.
[8] Normalization is a preprocessing step in the clustering technique, which is automatically done in the Python package tslearn.

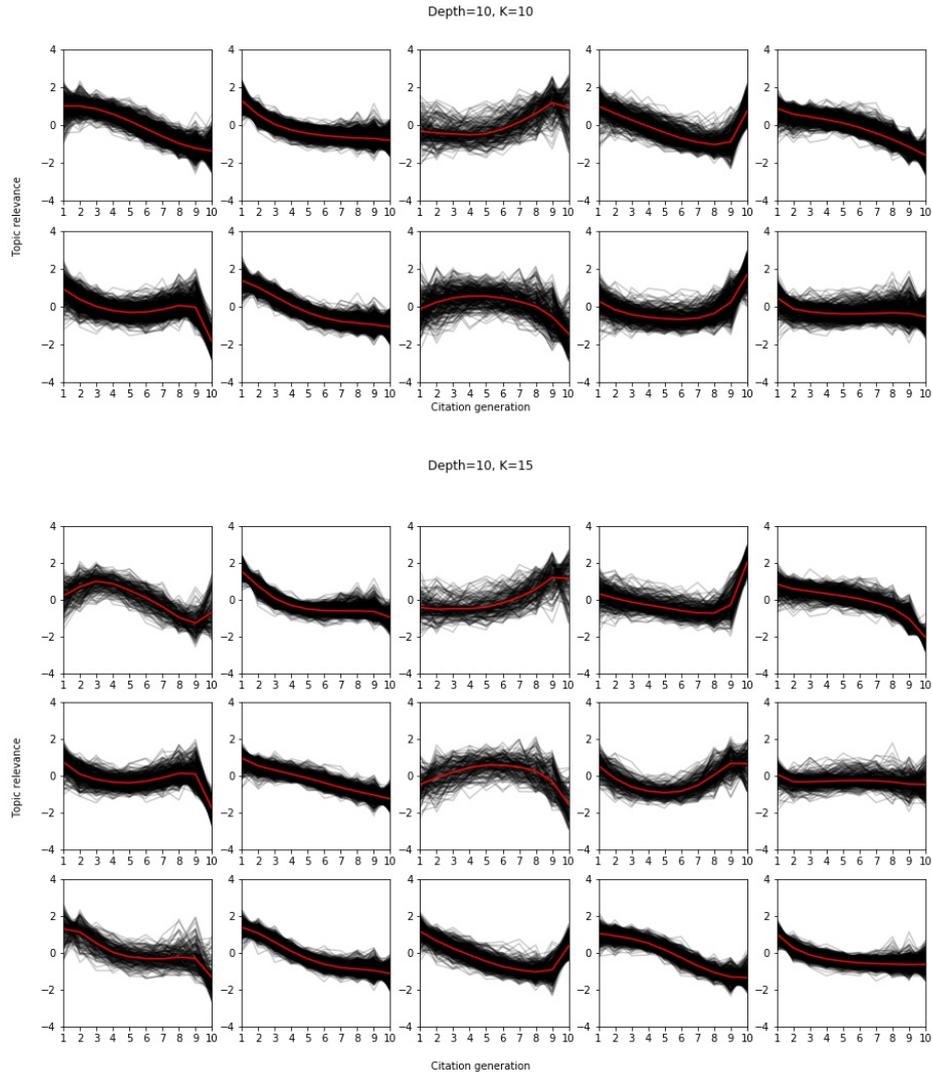

Figure 6. Typical patterns of relevance curves at depth = 10 (7,504 cascades, K = 10, 15)

For example, in citation cascades with a depth of 10 (Figure 6), we observe the following relevance-curve patterns:

(1) Decreasing mode. The relevance curve shows an overall decreasing trend with slightly different shapes. Some curves have a fast decrease and stabilize at a low relevance level; others decrease slowly throughout; and still others have a second decrease after going down to an already low level. Most relevance curves fit into one of these patterns.

(2) Increasing mode. A small number of relevance curves show an interesting increasing trend in which higher-order citations, not lower-order ones, are more relevant to the root article. In these curves, the increasing trend continues for several generations.

(3) Concave mode. The relevance declines to a minimum over several generations, then begins to rebound. The highest point post-rebound is sometimes lower, sometimes higher than the previous highest point; in either case, the curve forms a concave arc.

(4) Convex mode. Some curves briefly increase in relevance and then begin to

decrease; others decrease after a longer period of increase. These curves appear as a convex arc.

(5) Mixed mode. Some curves have relatively complicated shapes in which both concave and convex segments exist. Figure 6 shows that on such mixed-mode curves, the concave part usually appears prior to the convex part.

We have not observed a monotonic increasing mode in cascades with a depth of 20, 30 or 50: that is, in no cascade does the topic relevance keep growing for more than 20 generations. Yet we do observe "two-stage decrease" patterns in some curves with a depth of 20: in these, topic relevance stabilizes after decreasing to a low level, but then goes on to decrease to a lower level. For those cascades with a larger depth (e.g., 30 and 50), the relevance curves present more wave-shaped variations. Increasing portions or spikes are common when deep citation generations are reached.

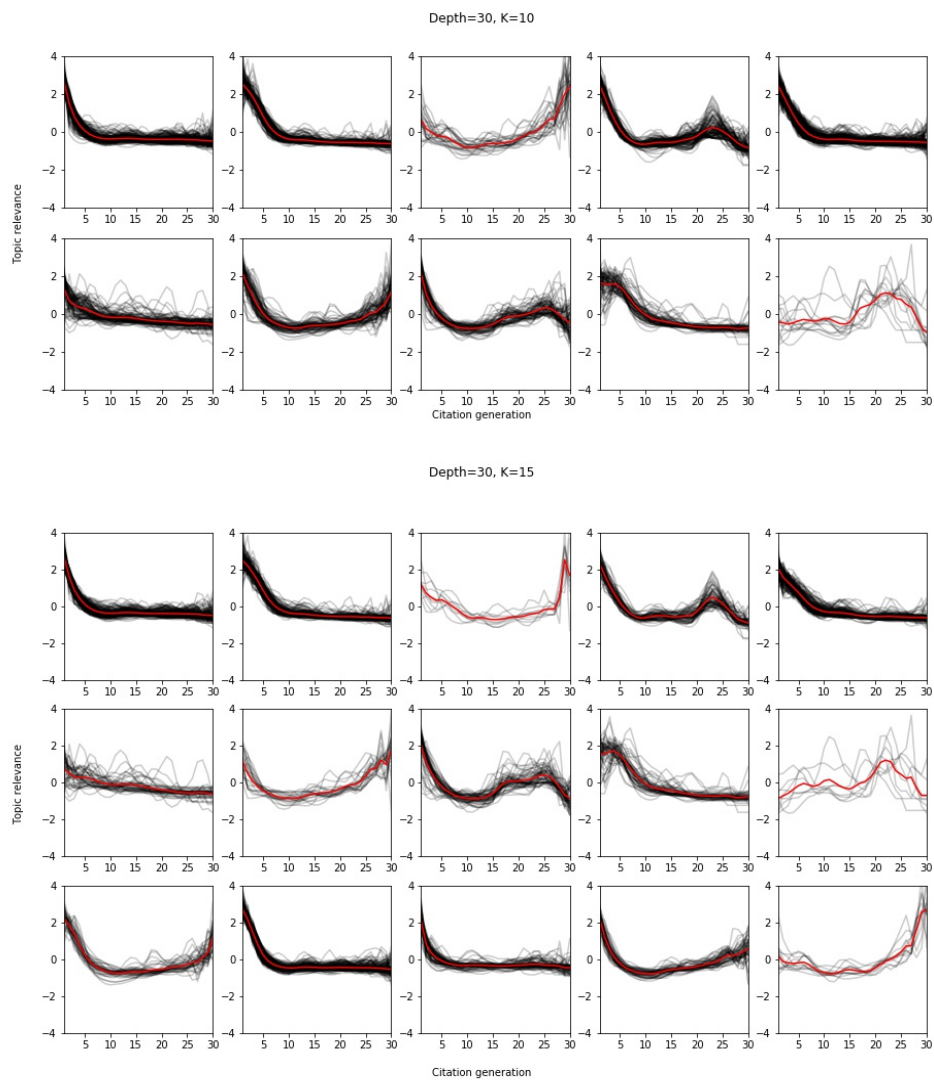

Figure 7. Typical patterns of relevance curves at depth = 30 (1,409 cascades, K= 10, 15)

## Relevance evolution across citation generations

In the previous section, we investigated topic relevance on the micro level (within an individual cascade). Here, we show how topic relevance changes across generations within the entire APS dataset. To be more specific, we first average the topic relevance for an article's citing articles in a certain citation generation, and then average the prior result for all root articles in the dataset. The overall topic relevance by generation is graphed in Figure 8. For comparison, we construct a random network, perform the same calculation, and plot its relevance curve[9]. The random network is constructed by reshuffling the APS citation network while preserving the out-degree, in-degree, and publication year for each node (Uzzi et al., 2013). The result is enlightening and intuitive, but also rewards a more detailed examination. Over the entire dataset, topic relevance shows a generally decreasing trend. It is slightly greater than 0.5 in the first citation generation and starts to decrease from the second citation generation onward, declining to less than 0.1 near the eighth generation. Topic relevance in the real citation network is distinctly higher than the random level up until the tenth generation. After that, relevance stabilizes at approximately the random level. Although slow decreases and fluctuations exist, we tend to consider them as natural variations, since the deviation from the random-network curve is very small. This indicates that topic relevance is still higher than random level in the second to tenth citation generations, especially in the second, third, and fourth generations ($> 0.25$).

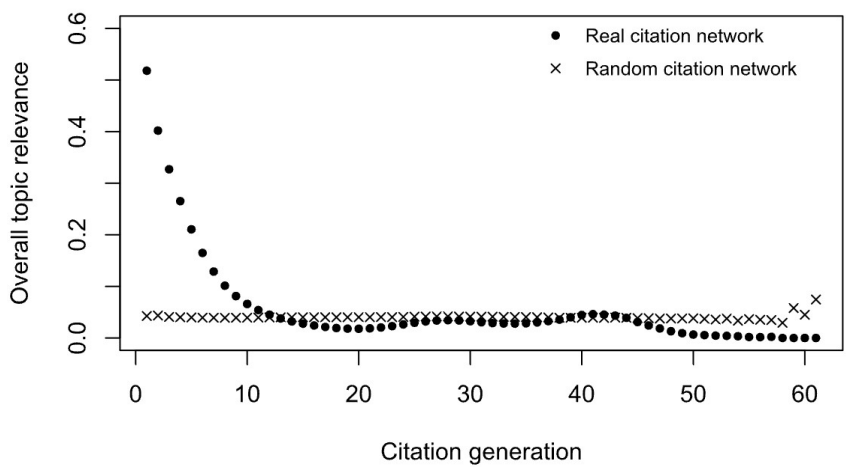

Figure 8. Evolution of overall topic relevance along citation generations

To test whether the findings above can be generalized regardless of other bibliometric factors, we graph overall topic relevance curves for different groups of papers based on publication type, journal, and research field[10], respectively. The results show very similar trends to those in Figure 8, particularly within early generations. Publication type is found to have very limited influence on the trends of the curves, as the five curves nearly overlap within this group. The only exceptions are several rises and fluctuations in the tails of some curves. These appear in the journals *Physical Review C* (PRC) & *Physical Review E* (PRE), and research fields PACS 1 & PACS 2. Interestingly, PRC and PACS 2 cover exactly the same field (nuclear physics), while PRE and PACS 1 also share similar fields, including statistical, nonlinear, and biological physics. This suggests that there does exist a field difference in the overall topic relevance curve. However, in view of the consistency between other curves and Figure 8, we tend to think that the rises and fluctuations are caused by changes within the two research fields themselves.

---

[9] Topic relevance in the random network is consistently low (roughly 0.03–0.05) across citation generations. The slight increase around the 60th generation is caused by the limited number of successors that can reach such high citation orders.

[10] There are five publication types (article, brief, rapid, comments, and reply), six journals (PRA, PRB, PRC, PRD, PRE, and PRL), and ten research fields corresponding to PACS first-level fields.

**Topic relevance of the first citation generation**

Figure 8 shows that the first-generation citing publications are most closely related to the root publication. Here we investigate the relationship between a publication's direct citation impacts and the topic relevance of its first-generation citations. Both total and average topic relevance are taken into consideration for the first citation generation.

Assuming that a related citation to an article is a reward or recognition of that article, we can estimate its overall direct recognition by adding up the relevance values of the first-generation citations. This is termed total relevance of the first citation generation. Figure 9 shows that total relevance (vertical line in the box plot) shows a strong increasing trend with the number of first-generation citations. Namely, little-cited articles have low total relevance in the first citation generation, whereas highly cited articles have high total relevance in the first citation generation. To put it another way, highly cited articles receive citations that are also related to themselves, resulting in the accumulation of total relevance. This reveals an aspect of the relation between scientific impact and scientific relevance: a scientific work has to gain quite a lot of recognition from subsequent works in domestic or related fields to achieve high impact.

Figure 10 shows, however, that average relevance exhibits an obvious decreasing trend with the total number of first-generation citations. That is, little-cited articles have high average relevance to their first-generation citations, and highly cited articles have low average relevance to their first-generation citations. For articles with more than 1,000 citations especially, the average relevance of first-generation citations is significantly lower than for articles with fewer citation counts. The low relevance for those highly cited articles obviously stems from being cited by more distantly related or even (seemingly) unrelated works. This exposes another aspect of the relationship between scientific impact and topic relevance: a scientific work has to inspire subsequent works from many other disciplines, particularly transdisciplinary works, to achieve the highest impact. Quite possibly, such scientific works distinguish themselves not only by their great novelty and originality, but by their broad applicability.

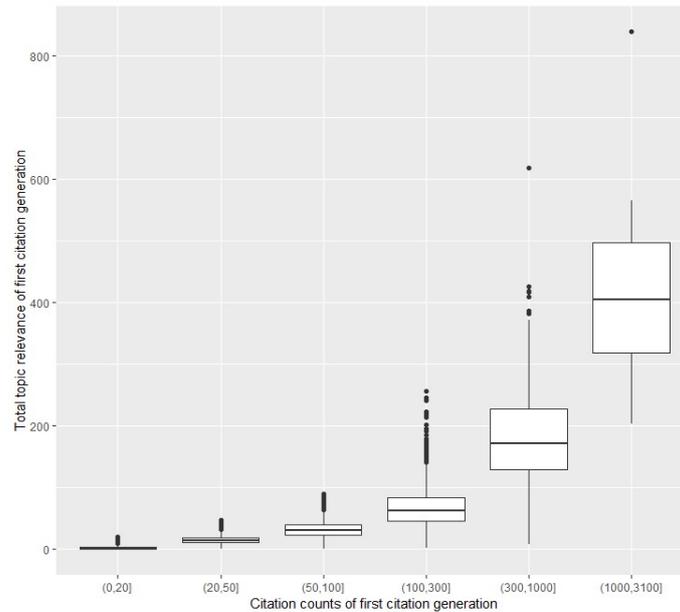

Figure 9. Total relevance increases with citation counts in the first generation

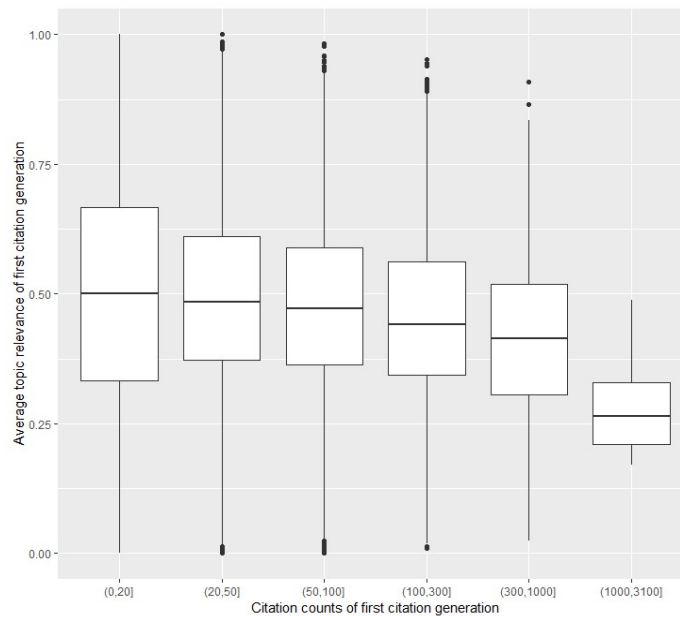

Figure 10. Average relevance decreases with citation counts in the first generation

## Citation cascade and direct impact

The direct impact of a scientific work is often reflected in the number of first-generation citations. Although direct impact may be influenced by many factors, we will next probe into the relationship between a citation cascade's structural properties (depth, width, size, structural virality, and topic relevance of both direct and indirect citations) and the root article's direct citation impact.

Figure 11 suggests an inverted U relation between an article's direct citation count and the depth of citation cascade. The depth is divided into seven segments. With increasing cascade depth, the median and quartile values of direct citation count for articles in each segment first increase and then decrease. This indicates that (1) the growth of citation cascade depth at first depends on the number of first-generation citations of the source article, since young cascades have more opportunities to grow when they have more direct citations; but (2) the dependence gets weaker and weaker when cascade depth reaches a threshold, after which the cascade's growth benefits more from indirect citations. Yet, as is shown in cascades with depth range (54, 63], extremely deep cascades still benefit from a high direct citation count. Figure 12 suggests a similar inverted U shape between direct citation count and structural virality of the corresponding cascade. The median and quantiles of direct citations first increase, then decrease with increases in structural virality. The result is reasonable, since we calculate structural virality by the average shortest path length between the root of a cascade and its children.

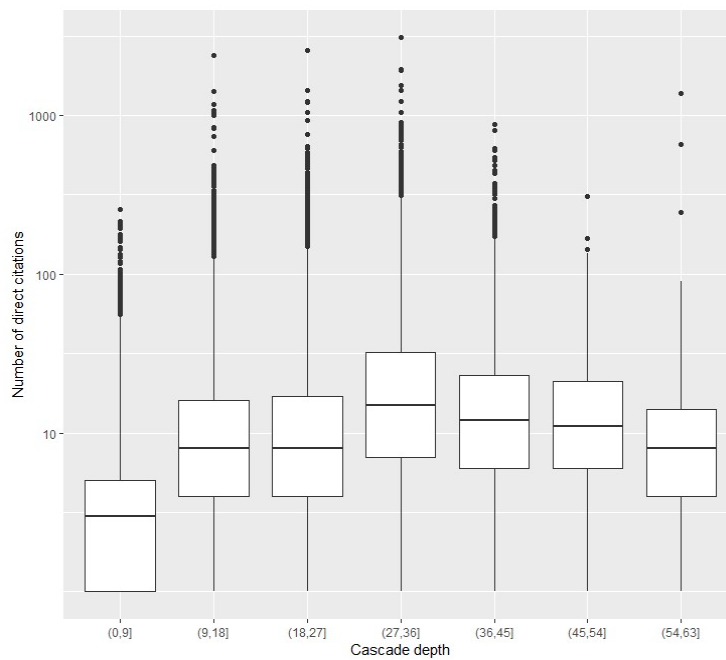

Figure 11. Number of direct citations increases and then decreases with cascade depth

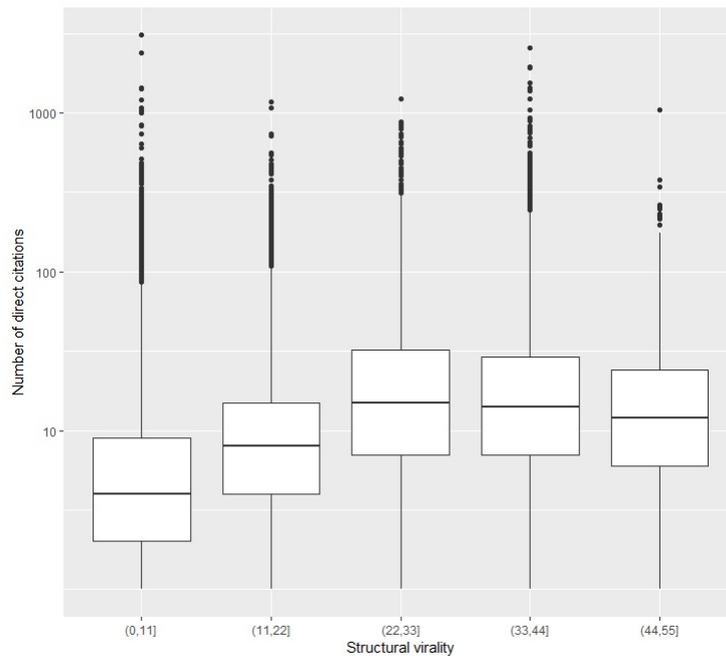

Figure 12. Number of direct citations increases and then decreases with structural virality

Previously, we analyzed the relation between direct citation counts and topic relevance, finding that total relevance increases with direct citation counts while average relevance simultaneously decreases. Now, we investigate how direct citation impact changes with average topic relevance by diving into the first three citation generations.

As is shown in Figures 13–15, the relation between (average) topic relevance and direct citation counts of source article presents an inverted U shape in all three citation generations. The number of direct citations increases when topic relevance increases in low-value ranges; however, it begins to decrease as topic relevance continues to rise. This observation indicates that it may not always be beneficial, in terms of overall impact, to be cited by a large proportion of articles on closely related topics, if the source article aims to achieve scientific impact at a high level. In the course of collecting scientific impact, being cited by related works can surely help an article make advances at lower levels; yet to achieve larger impact, its scientific contents must accordingly be appealing to and adopted by more works from "less related" research fields. The underlying mechanism reflected in this process is the scientific work's scalability, profundity and applicability to its successors. In this sense, scientific ideas that achieve high-level impact are often fundamental innovations that break the existing scientific structure and open up new research space; thus, they possess significant levels of scientific novelty.

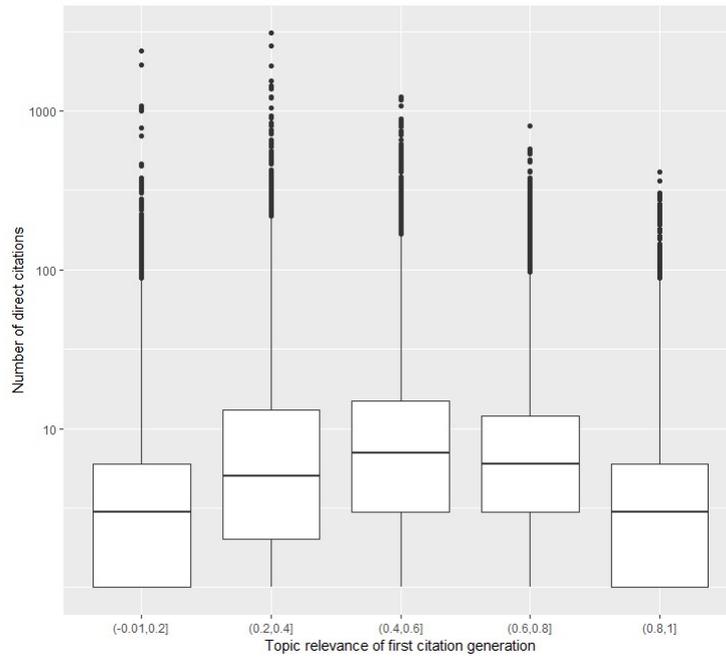

Figure 13. Number of direct citations increases and then decreases with topic relevance of the first citation generation

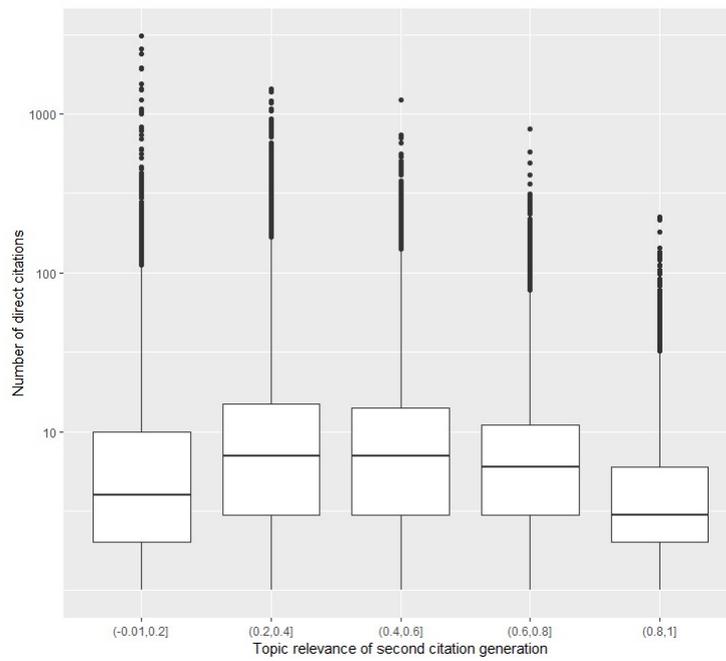

Figure 14. Number of direct citations increases and then decreases with topic relevance of the second citation generation

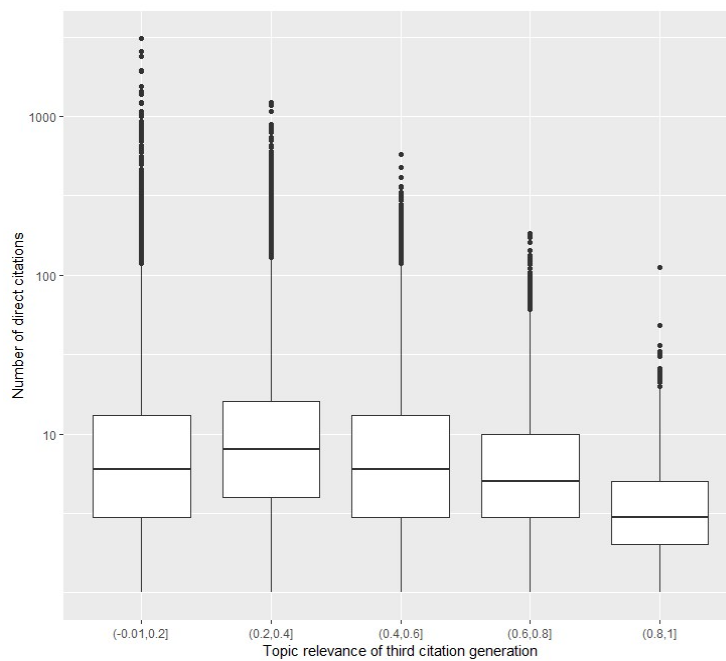

Figure 15. Number of direct citations increases and then decreases with topic relevance of the third citation generation

# Discussion and conclusions

## Citation cascade: a data structure for measuring high-order impact

Scientific literature often serves the function of recording and disseminating scientific knowledge. It follows that much of a work's value is displayed in its interaction with subsequent scientific works. Researchers and practitioners in scientometrics have traditionally given much attention to direct citations of scientific works, but have been less attentive to the issue of indirect or high-order citations. There are many reasons for this relative neglect, such as historical limitations in the accumulation of bibliometric data and construction of citation databases, insufficiency of computing power, and a dearth of theoretical tools. However, previous studies do demonstrate both the existence and the significance of indirect or high-order scientific impact (Rosvall et al., 2015). We propose the concept of the citation cascade as an approach for analyzing the evolution and continuation of scientific impact, as well as a dimension for examining the inheritance and mutation of scientific knowledge from a structural perspective. Herein, we present a large-scale study on a real dataset in physics to reveal the structural characteristics of citation cascade and to point out some practical implications of those features.

A citation cascade consists of a series of citing events directly or indirectly initiated by a single scientific publication, called the root publication. Mathematically, the cascade can be modeled as a directed acyclic graph stemming from the root publication. In this graph, the direction represents the flow of scientific knowledge, which is the same as the direction of being cited and thus the direction of the timeline. Apart from the root publication, a citation cascade contains one or multiple citation generations. The set of citing publications that are N steps from the root publication constitutes the Nth citation generation of the root publication. It is possible for a citing publication to exist in multiple citation generations simultaneously, and for citation relations to span different citation generations of the root publication. The structure of a citation cascade can be measured by such aspects as depth, width, structural virality, size and topic relevance. Citation cascades reflect not only the inheritance and extension of scientific knowledge, but also its mutation and development. Such dynamic relations build connections both within and beyond scientific fields even as they capture these connections in network form. Together, these relations reflect the inner structure of scientific development.

The analogy between cascades in scientific citations and in social media is by no means perfect, as their data structures are different. First, a Facebook or Twitter cascade often spreads simple content (e.g., a short message), whereas a citation cascade spreads scientific knowledge that is more complex. Second, whereas the same content is spread in an information cascade, a citation cascade disseminates multiple content items and generates new ones at the same time. Third, a child node is usually triggered by a unique parent node in an information cascade; in a citation cascade, the creation of a new child node depends on the integrated impact of all the parent nodes, and sometimes on such other factors as researchers' inspiration or serendipitous discoveries. Fourth, the main information content remains almost unchanged in such information cascades as retweets. However, mutation and development often occur in a citation cascade of scientific knowledge. Hence, we tend to say that an information cascade helps understand the mechanism underlying viral ideas, products, etc., while a citation cascade helps record the growth and evolution of scientific knowledge.

## Characteristics and patterns of citation cascades

The experimental results answer our previously proposed question: citation behavior in deep citation space is not totally random and can still be associated with the original publications. The width of citation cascades presents various shapes, generally showing a first-increase-and-then-decrease trend when charted against citation generation. Both the depth and size of citation cascades display a long-tailed distribution. Compared with a publication's direct citation counts, its overall citation cascade grows more rapidly and obviously over time. The impact of time on structural virality is twofold (Min, Sun & Ding, 2017). On the one hand, structural virality gradually grows with time; on the other hand, it will reach a limit where growth stops, while younger cascades retain the potential to grow further.

An interesting observation that needs further discussion is that the middle citation generations always have a relatively large number of papers (e.g. in Figure 3). Why does there exist such an up-and-then-down pattern? We propose an explanation for further discussion: two mechanisms are operating simultaneously. First, because a paper in a low-order generation is often cited by multiple (even dozens of) papers that are in a high-order generation, it is easily inferred that high-order generations tend to have more papers than low-order generations. We call this Mechanism 1: the deeper the citation generation, the more papers. Second, a paper in a low-order generation is usually older than a paper in a high-order generation (although this is not true in all cases). Older papers have a longer time to accrue citations than younger papers do, as the latter need more time to be cited. From this perspective, a paper in a low-order generation is likely to be cited by more papers. We call this Mechanism 2: the shallower the citation generation, the more papers. The two mechanisms are taking effect at the same time, leading the middle generations to have more papers.

Topic relevance intuitively reflects the inheritance and mutation of knowledge within citation cascades, revealing interesting characteristics in the APS dataset. It does not always reach its maximum in the first citation generation and then decrease afterward. Rather, the maximum value may appear in later citation generations. Topic relevance thus does not simply monotonically decrease with citation generation, but displays various forms in different cascades: decreasing, increasing, concave, convex, and mixed. As the depth increases, irregular curves with wavy shapes may appear, and topic relevance may rise again in high-order citations. Overall, topic relevance tends to undergo a decrease in the first few citation generations, then stabilize to the level expected of a random network after 10 generations. Two enlightening figures are worth mentioning in this connection: citations within four generations still have considerable relevance to the root publication, but topic relevance after ten generations becomes very weak. Our observations concerning citation cascades coincide with Christakis & Fowler's (2009) theory of "three degrees of impact," suggesting that a paper citation network is more like a network of impact than a network of connection (Travers & Milgram, 1969), in terms of research topic relevance. Apart from topic inheritance, topic mutation also sheds light on the development and evolution of scientific knowledge. We find that total relevance increases but average relevance decreases for papers receiving different numbers of citations. This difference occurs because total relevance always grows when a paper collects new citations and new relevance values are added. However, if the paper has a broad impact and thus continues to attract citations from foreign fields (where the citing publication might be less relevant), the average relevance value will be pulled down. This phenomenon reveals two complementary rules of scientific research: in order to achieve high impact, a scientific discovery needs not only to receive sufficient recognition from subsequent works in its domestic field, but also, perhaps more importantly, to inspire following works that transcend disciplinary boundaries. Such works, anchored in one field but with wide influence in other fields, are highly novel and broadly applicable.

The lengthwise growth of a citation cascade depends at first on direct citations to the root publication, but its further growth benefits more from indirect citations as cascade depth continues to increase. The direct influence of the root publication is maximized when the relevance between (direct or indirect) citing publications and the root publication is neither very high nor very low, but rather, down the middle (approximately (0.2, 0.4]). That is to say, being cited by merely "relevant" publications can only enhance impact at a low level; high-level gains in impact require a scientific work to inspire subsequent research in "less relevant" or "possibly relevant" fields.

## Implications for science evaluation

Direct citation has been a focus in science evaluation for a long time. On the basis of previous studies (Fragkiadaki et al., 2011; Hu et al., 2011; Rousseau, 1987), we provide more evidences of indirect (or high-order) citation impact. This may also help explain the existence of "under-cited influential publications" (Hu & Rousseau, 2016). We advocate the awareness of such high-order impact and suggest that indirect and cross-generation impact be considered in science evaluation. The number of citation generations to be included is an important issue in practice. The answer to this question may vary in different scenarios of application, but we find citations within four generations are overall substantially related to the root publication, echoing the speculations of Rousseau (1987). There seems to be no need to consider citations beyond ten generations, as topic relevance at this depth drops to a background level. We therefore suggest the proper inclusion of high-order citations (e.g., second to fourth generations) in related tasks of scientific impact evaluation.

However, the utilization of high-order citations is retrospective in nature and thus not suitable for contemporary analysis, as the accumulation of high-order citations takes time. Therefore, it is more appropriate to use this approach in the analysis and modeling of the history of science, where large amounts of historical data are available. In the historical development of science, for example, some publications that include foundational discoveries didn't receive many citations themselves but triggered a series of innovative works that became heavily cited. These publications should have held important positions in the history of science, but they are easily overlooked under traditional evaluation indicators based on direct citation. High-order citation analysis is naturally suitable for identifying such under-cited but foundational publications, thus helping uncover the true face and objective patterns of science development.

Moreover, the general approach of citation generation analysis, supplemented by the investigation of research topics, objects, methods and researchers across different generations, is beneficial to the delineation of the improvement and evolution of scientific knowledge. Traditionally, the analysis of the developmental history of a certain scientific field is often based on publication data retrieved from keyword search results or specific publication venues. As either the data lack semantic associations or a transdisciplinary perspective is missing, this approach inevitably limits the scope of analysis. Citation generation analysis takes advantage of the citation linkages across different generations of knowledge; thus, it has the virtues of including semantically associated knowledge components and of being open to both domestic and foreign scientific disciplines. This is also beneficial to interdisciplinary research in the developmental history of science.

## Limitations and future research

As the findings of the present study come only from the discipline of physics, we cannot assume that they fully apply to other disciplines. Further research is needed to verify the findings in other disciplines and datasets—for example, whether the unusual gap in Figure 5 will appear again, whether topic relevance will rise in high orders in other disciplines, and what the corresponding cause might be in that case. Citation cascades should also be used with caution in practice. The data structure can easily become very large, with a rapid concomitant rise in computational expense. We assume it is enough to consider the first several generations and that there is no need to go deeper. An interesting perspective for further study is topic relevance to the root publication. If high-order citations remain equally (or even increasingly) relevant to the root publication, does this type of citation cascade record a special process of scientific discovery? Does such a pattern indicate that subsequent works continue to follow and extend the research of the root publication? This deserves further study and deliberation by researchers and practitioners in science evaluation. Another enlightening perspective is (as one reviewer suggested) studying references and references of references, or backward generations (Hu et al., 2011). How many reference generations actually influence the research in the target article would be an interesting question to address in the future.

## Acknowledgments

This research was supported by the National Natural Science Foundation of China (NSFC No. 71904081, No. 71874077), Chinese Education Department Research Foundation for Humanities and Social Sciences (No. 19YJC870017), Jiangsu Province Social Sciences Foundation (No. 18TQC005) and the China Scholarship Council. The authors are grateful to the editor, reviewers , and Qing Ke from Northeastern University for helpful comments.